\documentclass[aps,prl,twocolumn,superscriptaddress,showpacs]{revtex4}
\usepackage{epsfig,graphicx,times}
\begin{document}
\title{Spin and Conductance-Peak-Spacing Distributions in 
       Large Quantum Dots:\\ A Density Functional Theory Study}

\author{Hong Jiang}
\affiliation{Department of Chemistry, Duke University, Durham, North Carolina 27708-0354}
\affiliation{Department of Physics, Duke University, Durham, North Carolina 27708-0305}
\affiliation{College of Chemistry and Molecular Engineering, Peking University, Beijing, China 100871}
\author{Harold U. Baranger}
\email[]{baranger@phy.duke.edu}
\affiliation{Department of Physics, Duke University, Durham, North Carolina 27708-0305}
\author{Weitao Yang}
\email[]{weitao.yang@duke.edu}
\affiliation{Department of Chemistry, Duke University, Durham, North Carolina 27708-0354}

\date{\today}
\begin{abstract}
We use spin-density-functional theory to study the spacing between conductance peaks and the
ground-state spin of 2D model quantum dots with up to 200 electrons. Distributions for different
ranges of electron number are obtained in both symmetric and asymmetric potentials. The even/odd
effect is pronounced for small symmetric dots but vanishes for large asymmetric ones, suggesting
substantially stronger interaction effects than expected.  The fraction of high-spin ground
states is remarkably large. 
\end{abstract}
\pacs{73.23.Hk, 73.40.Gk, 73.63.Kv}
\maketitle

The interplay of quantum mechanical interference and electron-electron interactions is a current
theme in many areas of solid-state physics: the 2D metal-insulator transition, interaction
corrections in mesoscopic systems, and efforts toward solid-state quantum computing, for instance.
A semiconductor quantum dot (QD) \cite{Kouwenhoven97,Alhassid00RMP} -- a nano-device in which electron motion
is quantized in all three dimensions -- is a particularly simple system in which to study this
interplay. In Coulomb blockade experiments in the electron tunneling regime, the conductance
through the dot varies strongly as a function of gate voltage, forming a series of sharp peaks.
For closed dots at low temperature, both the positions and heights of the peaks encode information
about the dot's ground state. In particular, the spacing between adjacent conductance peaks
is proportional to the second difference of the ground state energy with respect to electron
number $N$, $\Delta_2 E(N) \!\equiv\! E_{gs}(N+1)\!+\!E_{gs}(N-1)\!-\!2 E_{gs}(N)$, which is
often called the addition energy. Furthermore, the ground state spin of the QD can be inferred
from the shift in position of the conductance peaks upon applying a magnetic field.

The addition energy varies because of changing interference conditions either as $N$
changes or from dot to dot, leading to a conductance-peak-spacing distribution.  Previous
theoretical work addressing this distribution can be divided into roughly two types: First,
computational approaches addressed small dots with randomly disordered potentials -- both exact
diagonalization \cite{Sivan96,Prus96,Berkovits98} and self-consistent field methods (Hartree-Fock
\cite{Cohen99,Walker99,Levit99,Ahn99,Bonci99} or density functional theory \cite{Stopa96,Hirose02}).
Second, a semi-analytic treatment of large dots was developed based on general statistical
assumptions \cite{Blanter97,Alhassid00RMP,Ullmo01b,Usaj01,Usaj02,Aleiner02}:  An important contribution to
the variation comes from the single-particle energy; to treat this for irregular quantum dots,
one assumes that the single-particle dynamics is classically chaotic, and so the single-particle
quantum properties can be described by random matrix theory (RMT) \cite{Bohigas90,Alhassid00RMP}.
A random-phase approximation (RPA) treatment of the screened electron-electron interaction was then combined with such an
RMT description of single-particle states to describe dots with large $N$.

The results of these two approaches are quite different. First, for the zero temperature
peak-spacing, the small dot calculations yield Gaussian-like distributions while the large $N$
results are non-Gaussian. Second, spin degeneracy causes a significant ``even/odd effect'' in
the large dot approach: the distribution for $N$ even is very different from that for $N$ odd.
Third, with regard to the ground state spin \cite{Brouwer99,Kurland00,Jacquod01,Hirose02,Oreg01},
the small $N$ calculations find an enhancement of the low spin states compared to the large
QD approach.

Experimental work to date \cite{Sivan96,Simmel97,Patel98,Luscher01,Ong01} has unfortunately
failed to probe the ground state addition energy distribution or ground state spin of generic
systems. In the more recent experiments, either temperature obscured the ground state properties
\cite{Patel98,Ong01,Usaj01,Usaj02} or the dot was regular in shape \cite{Luscher01}.

Our aim here is to bridge the gap between the two theoretical approaches and in so doing highlight
the need for more experiments.  We have used the Kohn-Sham (KS) spin density-functional
theory (SDFT) \cite{ParrYang89} to study both the peak-spacing and the spin distribution for
2D model QDs.  Using an efficient algorithm
   \footnote{In our program, the wave functions are represented in real space, and the
kinetic energy operator is applied using a fast-sine transform. The Hartree potential
is calculated using Fourier convolution. The KS equations are solved by a direct
minimization preconditioned conjugate-gradient method \cite{Payne92}.
   },
we obtain statistics with $N$ up to 200.  This is the first calculation, as far as we know,
for large realistic QDs.

Our primary result is that the effective electron-electron interaction that emerges is
substantially stronger than that predicted from the RPA-RMT treatment. The evidence is two-fold:
(1)~the addition energy distribution is Gaussian-like with no discernible even/odd effect, and
(2)~the probability of having a high spin state ($S \ge 1$) is larger than the maximum possible
from RPA-RMT.

In SDFT, the energy of the interacting system is expressed as a functional
of the spin up ($\alpha$) and down ($\beta$) electron densities, $\rho_{\alpha}(\mathbf{r})$
and $\rho_{\beta}(\mathbf{r})$.  (Effective atomic units are used: for
GaAs-AlGaAs QDs, the values are 10.08 meV for energy and 10.95 nm for length.)  The functional
is built out of four parts. Two are trivial -- the response to the external potential $V_{ext}$
and the Coulomb energy.  The kinetic energy of a non-interacting reference system 
is included explicitly:
$T_s\lbrack
\rho_{\alpha},\rho_{\beta} \rbrack=\sum_{i,\sigma}
\langle \psi_{i\sigma}|-\frac{1}{2}\nabla^2|\psi_{i\sigma}\rangle $ 
with $\rho_{\sigma}= \sum_i |\psi_{i\sigma}|^2 $ and $\sigma=\alpha,\beta$. 
The final term is, of course, the exchange-correlation energy 
$E_{xc}\lbrack \rho_{\alpha},\rho_{\beta} \rbrack $.
Thus the KS-SDFT functional is
   \begin{eqnarray} 
E\lbrack \rho_{\alpha},\rho_{\beta}
\rbrack =T_s\lbrack \rho_{\alpha},\rho_{\beta}
\rbrack +\int V_{ext}(\mathbf{r}) \rho(\mathbf{r})d\mathbf{r} {} \nonumber\\
+\int\frac{\rho(\mathbf{r})\rho(\mathbf{r}')}{|\mathbf{r}-\mathbf{r}'|}d\mathbf{r}d\mathbf{r}'
+ E_{xc}\lbrack \rho_{\alpha},\rho_{\beta}
\rbrack.\qquad \label{eq:KS} 
   \end{eqnarray} 
The ground state energy is obtained by minimizing the functional with respect to the spin densities under
the constraints $\int \rho_\sigma(\mathbf{r})d\mathbf{r}=N_\sigma$.  We use the standard local
spin density approximation (LSDA) for $E_{xc}$ which works well in various semiconductors; in
particular, we use Tanatar and Ceperley's 2D parametrization \cite{Tanatar89}.  For 2D clean
QDs, comparisons with quantum Monte-Carlo calculations for $N \le 8$ and interaction strength
parameter $r_s \le 8.0$ have shown that LSDA works well for both the ground state spin and
energy \cite{Egger99,Pederiva00}.

We use a quartic potential to model 2D QD systems,
   \begin{equation}
V_{ext}(\mathbf{r})=a \lbrack \frac{x^4}{b}+b  y^4 -2 \lambda x^2 y^2 + \gamma (x^2 y - x
y^2)r \rbrack.
      \label{eq:qop}
   \end{equation}
Both the classical dynamics and the single-particle quantum mechanics at $\gamma=0$ have
been studied in detail \cite{Bohigas93}: the system evolves continuously from integrable to
fully chaotic as $\lambda$ changes from 0 to 1. The parameter $\gamma$ breaks the four-fold
symmetry. The prefactor is $a=1.0\times 10^{-4}$; this allows the electrons to spread so that
the interaction strength parameter, $r_s$, is about 1.5, close to experimental conditions.

\begin{figure}[t]
\includegraphics[width=3.0in]{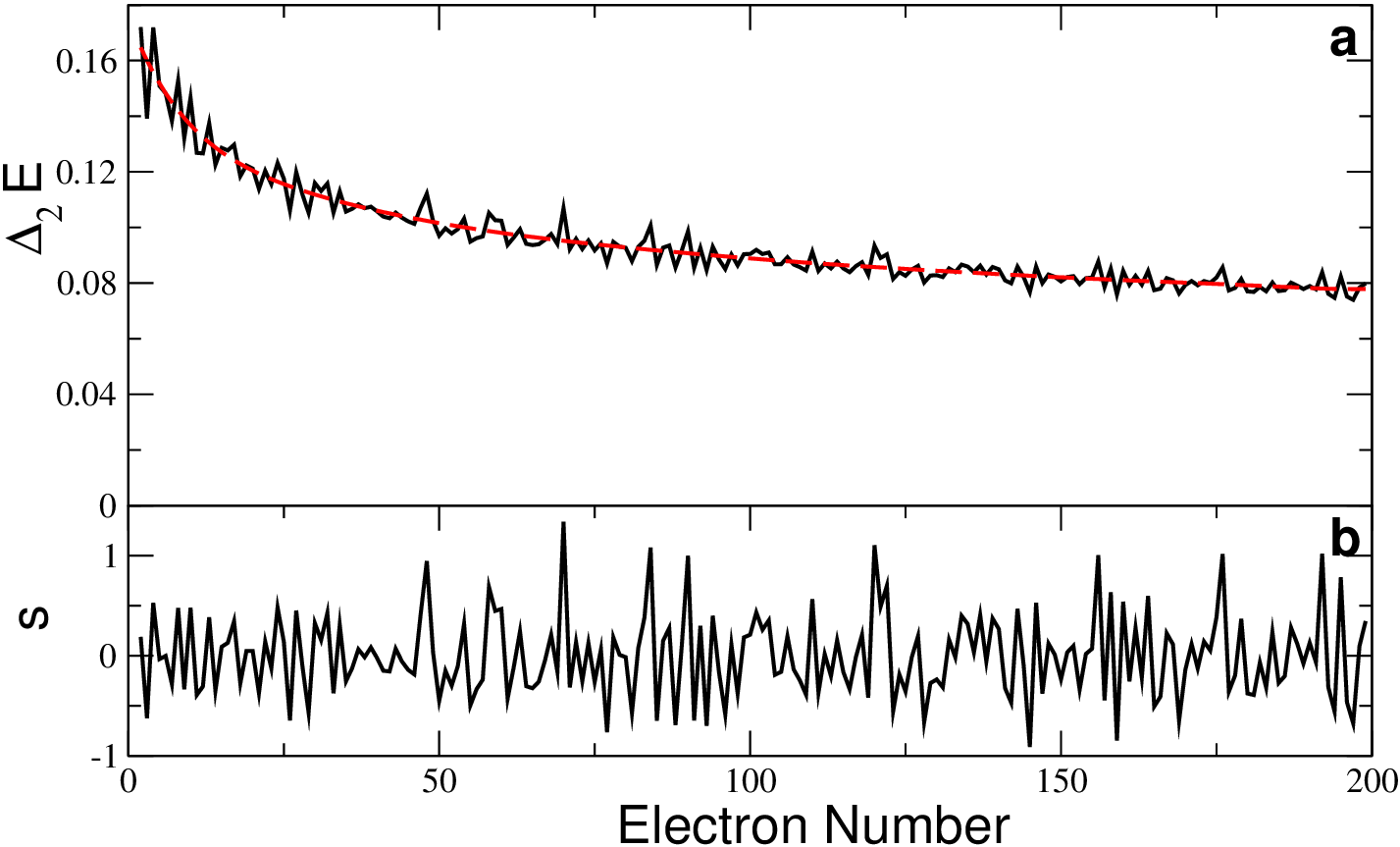}
\caption{\label{fig:data} Addition energy as a function of electron number. 
(a)~Raw addition energy data (solid) and its polynomial fit (dashed) to remove the change in
classical charging energy. 
(b)~Addition energies scaled by the mean level-spacing after removing smooth part.
($\lambda = 0.6$ and $\gamma = 0$.)}
\end{figure}

For a given $V_{ext}$, we calculate the total energy at several values of electron number $N$ and
total spin $S$. Selecting the minimum energy determines the ground state energy $E_{gs}$ and spin
$S_{gs}$ as a function of $N$.  The addition energy is then calculated as the second difference
of $E_{gs}(N)$; an example is shown in Fig.~\ref{fig:data}(a).  To obtain good statistics,
in both the symmetric and asymmetric cases we calculate five sets of data with different parameters
   \footnote{The parameter values used in our calculations are as follows: In all cases,
$a=1.0\times 10^{-4}$ and $b=\pi/4$. For the symmetric case $\gamma=0$, we take $\lambda
$ to be 0.53, 0.6, 0.67, 0.74, and 0.81. For the asymmetric case we choose five sets of
$(\lambda$,$\gamma)$: (0.53, 0.1), (0.565, 0.2), (0.6, 0.1), (0.635, 0.15), and (0.67, 0.1).
   }.
A correlation analysis shows that both the single-particle level spacing (SPLS) 
and addition energy from the different sets are statistically independent.

Since the slow decrease in  $\Delta_2 E (N)$ is a classical effect -- the increasing capacitance as
the dot becomes bigger -- we remove it by fitting a polynomial to find the smooth part $\langle
\Delta_2 E(N) \rangle$.  To compare with experiments, the addition energy is scaled by the
mean-level spacing found from the average electron density $n$ through $\Delta \!=\! 2 \pi \hbar^2
\langle n \rangle/m^* N$.  The resulting dimensionless spacing is denoted $s \!\equiv\! [\Delta_2
E(N) \!-\! \langle \Delta_2 E(N) \rangle]/ \Delta$; see Fig.~\ref{fig:data}(b).  Note that the
typical scale of $s$ is 1; that is, fluctuations of the addition energy are on the scale of
the single-particle mean-level spacing.

\begin{figure}
\includegraphics[width=3.3in]{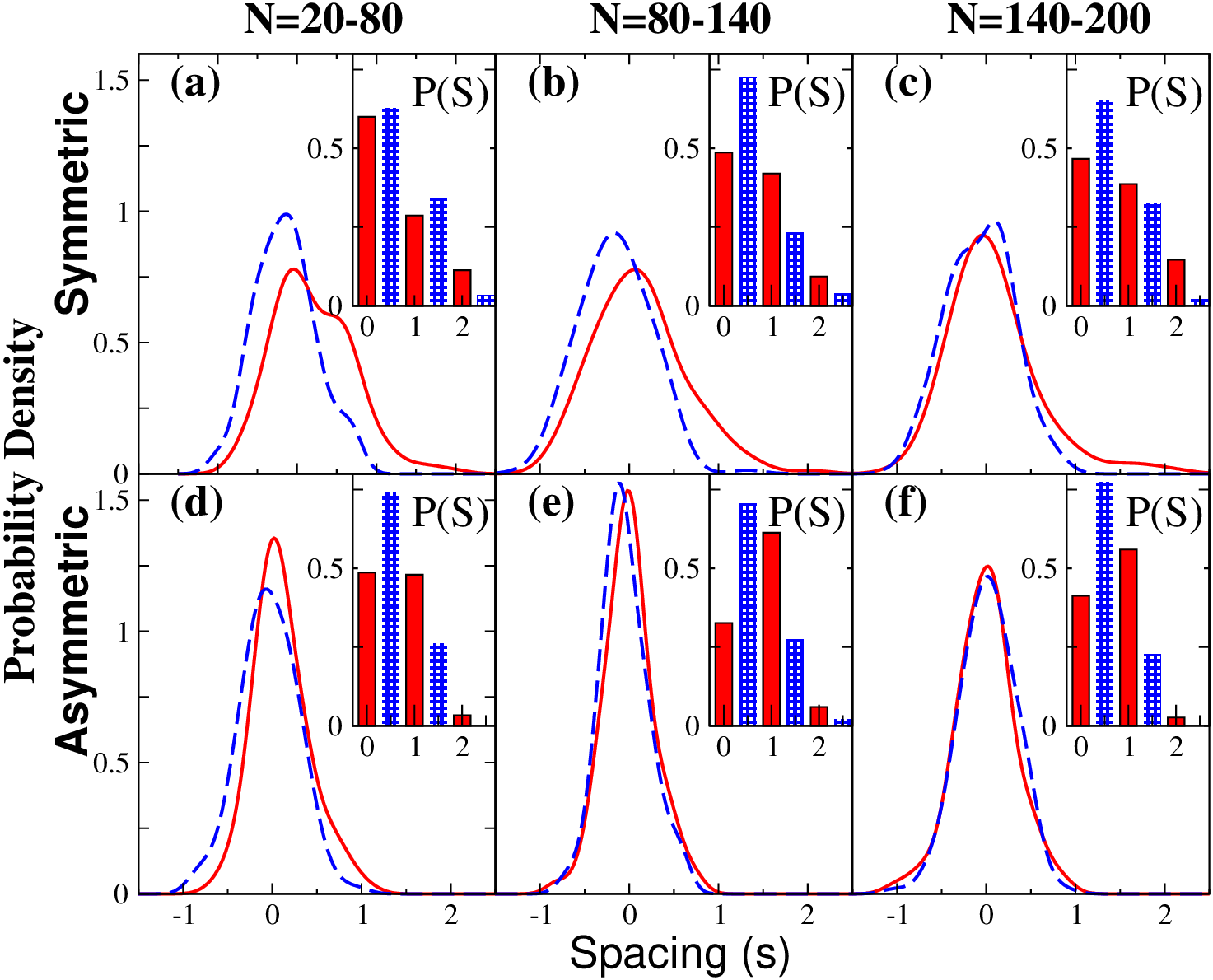}
\caption{\label{fig:distr} 
Distributions of peak-spacing, $s$, for even (solid) and odd (dashed) $N$.  The columns correspond
to the different ranges of electron number indicated, while the first row [(a)-(c)] is for a
symmetric external potential and the second for asymmetric [(d)-(f)].  The inset of each figure
shows the corresponding spin distribution for $N$ even (sold) and odd (skeleton).  Both the lack
of even/odd effect and the large spin in panel (f) is striking and indicates an unexpectedly large
effective electron-electron interaction.  (A sliding window is used in estimating the probability
density, yielding a smooth curve rather than a histogram; each curve is made from 150 data points 
using a Gaussian window of width 0.3.)}
\end{figure}

We find the distribution of $s$ for even and odd $N$ in three ranges of electron number, $N
\!=\! 20$-$80$, $80$-$140$, and $140$-$200$, in both symmetric and asymmetric external potentials.
(For the smallest, mean, and largest values of $N$, $r_s$ is 2.5, 1.6, and 1.4, respectively.)
The main features of the results, shown in Fig.~\ref{fig:distr}, are as follows.
   (1)~{\em Even/odd:} There is a difference in the distribution for $N$ even or odd in
the small $N$ range, and the parity effect in the symmetric case is more pronounced than for
asymmetric potentials.  But there is a striking {\em absence} of even/odd effect in the asymmetric
large $N$ case [panel (f)].
   (2)~{\em Shape:} The distributions are Gaussian-like. This extends the previous small $N$
results, disagrees with the RPA-RMT results for large $N$, and is consistent with experiments
\cite{Patel98,Ong01} (where, however, the ground state properties are obscured by temperature
effects \cite{Usaj01,Usaj02}).
   (3)~{\em Small/large:} The spacing distributions for small $N$ are different from
those for large $N$.  Since the experimental QDs generally involve tens to hundreds of
electrons, one must be cautious in generalizing to large dots conclusions drawn from studying
small dots.
   (4)~{\em Symmetric/Asymmetric:} Both the variance of the peak-spacing and the magnitude of the
even/odd effect is larger in the symmetric case.

The insets in Fig.~\ref{fig:distr} show histograms of the spin distribution for the corresponding
electron number ranges.
A remarkable feature is the significantly {\it higher} fraction of high spin ground states
that we find at large $N$ than in either previous SDFT investigations of small disordered
dots \cite{Hirose02} (for small $N$, we agree, of course, with previous results) or other
investigations. Especially in the asymmetric case, $P(S \!=\! 1)$ is even higher than $P(S
\!=\! 0)$ for $N \!=\! 80$-$200$!  Again we see a clear dependence on the electron number range:
while the spin distribution for odd $N$ changes little as $N$ increases, the spin distribution
for even $N$ is quite different for small and large $N$.  Comparing spin distributions in the
presence and absence of symmetry, we see a larger high-spin fraction in the asymmetric case
for all three ranges of $N$.

This last result is at first surprising: generally one expects increased symmetry to increase
the spin -- as in Hund's rule for the spin of atoms.  We can use random matrix theory, however,
to show that there are competing effects here -- namely the statistics of the eigenenergies
vs. the statistics of the eigenfunctions. For simplicity we consider the simplest RMT: a
two-electron two-level model in which we neglect spatial correlations beyond a wavelength.
In a Hartree-Fock framework and under the assumption that the orbitals remain the same for
different spin configurations, the energy difference between the singlet ($S \!=\! 0$) and
triplet ($S \!=\! 1$) states is
   \begin{equation}
\delta E\equiv E_{S=1}-E_{S=0}=\delta\varepsilon+(J_{12}-K_{12})-J_{11},
      \label{eq:two-level}
   \end{equation}
where $\delta\varepsilon$ is the single-particle level
spacing, $J_{ij}= \int d\mathbf{r}d\mathbf{r'} |\psi_i(\mathbf{r})|^2
v_{scr}(\mathbf{r},\mathbf{r}')|\psi_j(\mathbf{r})|^2$ are the Coulomb energies, and $K_{12}=\int
d\mathbf{r}d\mathbf{r'} \psi_1^*(\mathbf{r}) \psi_2^*(\mathbf{r}') v_{scr}(\mathbf{r},\mathbf{r}')
\psi_1(\mathbf{r}') \psi_2(\mathbf{r})$ is the exchange energy. Here instead of using the
bare Coulomb interaction, we use the screened potential $v_{scr}(\mathbf{r},\mathbf{r}')$
in order to implicitly account for the other electrons. Qualitatively, the screened
interaction is approximately zero-range, $v_{scr}(\mathbf{r},\mathbf{r}') \to (A \Delta/2)
\delta(\mathbf{r}-\mathbf{r}')$ where $\Delta$ is the mean single-particle level spacing
and $A$ is the area.  In this limit, the second term in Eq.~(\ref{eq:two-level}) vanishes,
and the third term is proportional to the ``inverse participation ratio'' (IPR) defined by $I
\!\equiv\! A\int d\mathbf{r} |\psi(\mathbf{r})|^4 $.  Therefore, in the zero-range limit and
considering the time-reversibility of the system
   \footnote{
In the time-reversal symmetric case, there are three contributions to the average IPR, the
direct, exchange, and Cooper pairings. However, when using the IPR to estimate the magnitude
of interaction effects, one must keep in mind that higher-order processes in the screened
interaction may renormalize the magnitude. This is well-known for the Cooper channel; since
its final magnitude is small, it should be neglected. We take this into account by using $2I/3$
in the estimate of the interaction effect rather than the full $I$.
  }, we have
   \begin{equation}
\delta E_{\rm zero-range}=\delta \varepsilon - I\Delta/3 \,.
      \label{eq:E_zero-range}
   \end{equation}
There is clearly a competition here between the level spacing -- a large spacing tends to decrease
the spin -- and the statistics of the wave functions -- increased localization increases $I$
and leads to a larger spin.

\begin{figure}
\includegraphics[width=2.5in]{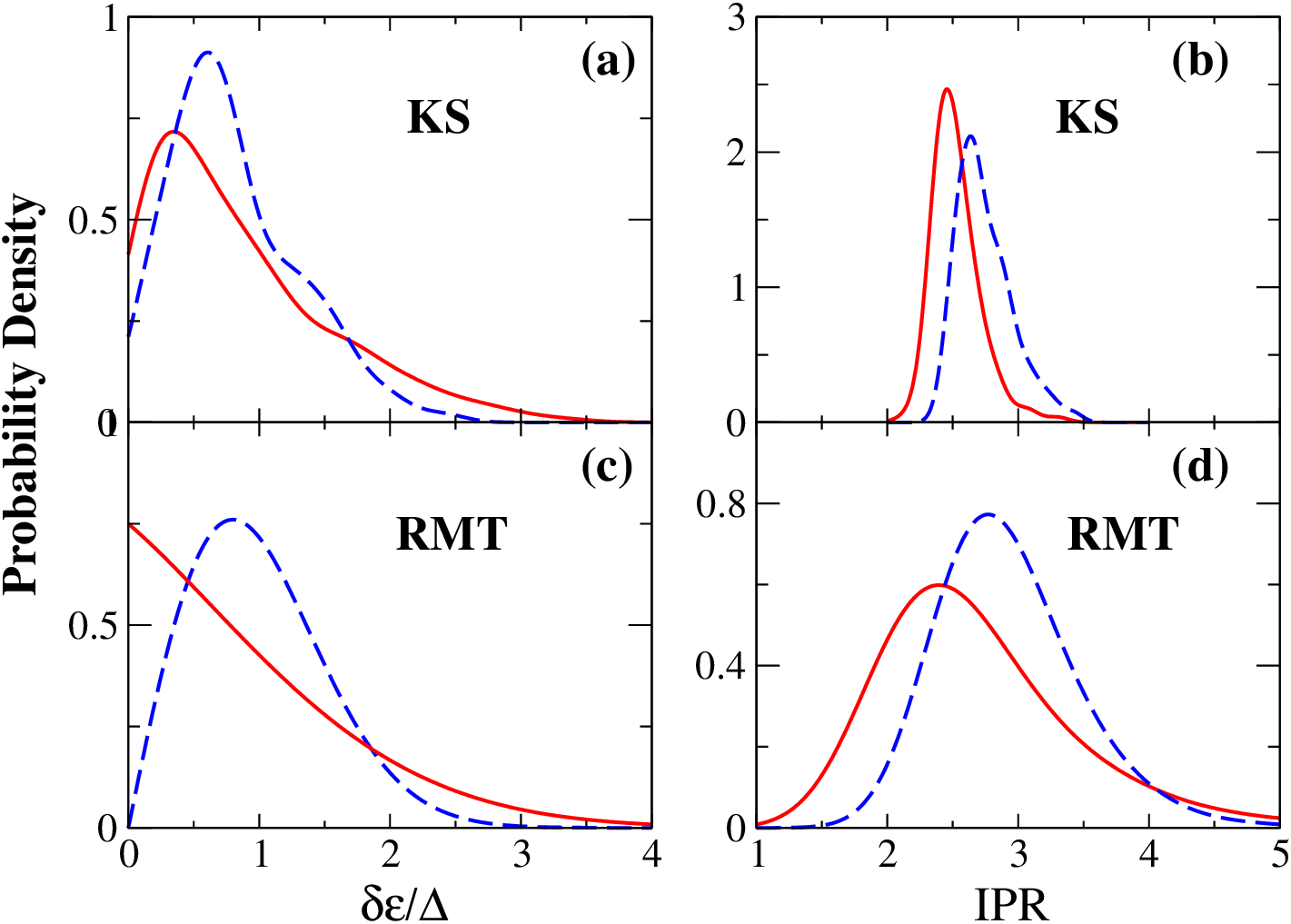}
   \caption{\label{fig:analysis} 
Distributions of the single-particle level spacing (scaled by the mean) and the IPR
from Kohn-Sham calculations and from random matrix theory.
   (a)~The distribution of SPLS calculated from top KS orbital energies for symmetric (solid)
and asymmetric (dashed) case respectively.
   (b)~The distribution of IPR calculated from top-level KS orbital wave functions for symmetric
(solid) and asymmetric (dashed) case respectively.
   (c)~The distributions of SPLS for a single GOE (dashed) and the
superposition of four GOE (solid) according to RMT.
   (d)~The distributions of IPR calculated from the eigen-vectors of real symmetric random matrix
with dimension $M \!=\! 20$ (solid) and $M \!=\! 80$ (dashed). }
\end{figure}

To see the competition explicitly, we calculate the distribution of the SPLS $\delta \varepsilon$
and that of the IPR from top-level KS orbital energies and wave functions in both the symmetric
and asymmetric cases. Results are shown in Fig.~\ref{fig:analysis}. While the symmetric case has
a higher probability of small level-spacing, the mean value of the IPR is also smaller in the
symmetric case.  Our overall result for the spin distribution -- the decrease in probability of
$S=1$ upon introducing symmetry [Fig. 2(f)] -- shows that the effect of wave function statistics
is stronger in our system.

The trend here is captured by the random matrix theory. For an asymmetric chaotic external
potential, the distribution of the SPLS is that of the Gaussian orthogonal ensemble (GOE). In
the symmetric chaotic case, however, the SPLS statistics is the superposition of four GOE's
\cite{Bohigas90}, one for each symmetry class. Fig.~\ref{fig:analysis}(c) shows these two
distributions.  Clearly, the superposition greatly reduces the nearest-neighbor level repulsion,
which implies that spatial symmetry favors a high-spin ground state, in accordance with Hund's
rule.

For the wave function statistics, RMT suggests that the single-particle wave functions
of classically chaotic systems are described by $M$-dimensional random unit vectors
\cite{Alhassid00RMP,Ullah64}.  The shape of the resulting distribution of IPR depends on $M$,
as shown in Fig.~\ref{fig:analysis}(d) for $M \!=\! 20$ and $80$. These values of $M$
are chosen to correspond to the number of independent orbital levels: approximately $N/2$
in the asymmetric case because of spin, and $N/8$ for the four-fold symmetric external potential.
Symmetry reduces the effective $M$ and so reduces the IPR, acting against a high-spin ground state.
Hence there is a competition between the SPLS and the IPR.

Note that the dependence of the IPR on $M$ also explains the change in the spin distribution
as electron number varies: The wave functions for small $N$ have a smaller effective $M$,
hence a smaller IPR. Thus a low-spin ground state is more favorable. Similarly, a three-level 
analysis provides a qualitative explanation of odd N results. 

The results in Fig.~\ref{fig:analysis} show, of course, that the calculated distributions of
both the SPLS and the IPR agree with RMT only qualitatively.  The fluctuation of the IPR is
particularly striking: it is much smaller for the KS wave functions than in RMT.  We believe
this is due to the neglect of spatial correlations in our very simple RMT.

In summary, by studying a model 2D quantum dot with up to $N \!=\! 200$ electrons, we have found
new phenomena. Both the statistics of ground state spin and the spacing between conductance
peaks depend on the electron number, as well as on the spatial symmetry. The results for
large electron number and asymmetric potential are surprising: the shape of the peak-spacing
distribution is Gaussian-like, the even/odd effect vanishes, and there is a substantial fraction
of large spin ground states ($S \!\ge\! 1$).  These effects imply a strong effective or residual
electron-electron interaction.

This is remarkable considering that conditions in our dots are not extreme: $r_s \!\sim\! 1.5$
corresponds to a moderate bare interaction strength, and the dimensionless conductance is
large, $g \!\sim\! 4$.  In fact, previous work using the RPA-RMT approach suggested that we
should obtain a strong even/odd effect \cite{Ullmo01b,Usaj02}. Conversely, to obtain the spin
and peak-spacing distributions that we find here from the RPA-RMT model requires an effective
exchange constant of $J_s \!\sim\! 0.6 \Delta$, larger than the maximum value possible in RPA.
The origin of this unexpectedly large residual interaction is not presently known, and we leave
it for future investigation.

We close with a caveat and a comment: First, this work is based on the 2D local spin density
approximation whose validity, though already verified for \textit{small} parabolic QD's
\cite{Egger99,Pederiva00}, is not well tested for \textit{large} non-parabolic quantum dots as
studied here.  Our result highlights the need to go beyond RPA-RMT and perform a real Fermi
liquid theory study.  On the other hand, note the recent experimental work in Ref.
\onlinecite{Folk01} which, though still preliminary because of insufficient data, indicates
a high probability of large-spin ground states. The surprisingly large effective interactions
found here suggest that more experiments should be a high priority.

We appreciate discussions with D. Ullmo and G. Usaj. This work was supported in part
by NSF Grant No. DMR-0103003 and the North Carolina Supercomputing Center.



\begin{thebibliography}{34}
\expandafter\ifx\csname natexlab\endcsname\relax\def\natexlab#1{#1}\fi
\expandafter\ifx\csname bibnamefont\endcsname\relax
  \def\bibnamefont#1{#1}\fi
\expandafter\ifx\csname bibfnamefont\endcsname\relax
  \def\bibfnamefont#1{#1}\fi
\expandafter\ifx\csname citenamefont\endcsname\relax
  \def\citenamefont#1{#1}\fi
\expandafter\ifx\csname url\endcsname\relax
  \def\url#1{\texttt{#1}}\fi
\expandafter\ifx\csname urlprefix\endcsname\relax\def\urlprefix{URL }\fi
\providecommand{\bibinfo}[2]{#2}
\providecommand{\eprint}[2][]{\url{#2}}

\bibitem[{\citenamefont{Kouwenhoven et~al.}(1997)\citenamefont{Kouwenhoven,
  Marcus, McEuen, Tarucha, Wetervelt, and Wingreen}}]{Kouwenhoven97}
\bibinfo{author}{\bibfnamefont{L.~P.} \bibnamefont{Kouwenhoven}},
  \bibinfo{author}{\bibfnamefont{C.~M.} \bibnamefont{Marcus}},
  \bibinfo{author}{\bibfnamefont{P.~L.} \bibnamefont{McEuen}},
  \bibinfo{author}{\bibfnamefont{S.}~\bibnamefont{Tarucha}},
  \bibinfo{author}{\bibfnamefont{R.~M.} \bibnamefont{Wetervelt}},
  \bibnamefont{and} \bibinfo{author}{\bibfnamefont{N.~S.}
  \bibnamefont{Wingreen}}, in \emph{\bibinfo{booktitle}{Mesoscopic electron
  transport}}, edited by \bibinfo{editor}{\bibfnamefont{L.~L.}
  \bibnamefont{Sohn}},
  \bibinfo{editor}{\bibfnamefont{G.}~\bibnamefont{Sch{\"o}n}},
  \bibnamefont{and} \bibinfo{editor}{\bibfnamefont{L.~P.}
  \bibnamefont{Kouwenhoven}} (\bibinfo{publisher}{Kluwer},
  \bibinfo{address}{Dordrecht}, \bibinfo{year}{1997}), pp.
  \bibinfo{pages}{105--214}.

\bibitem[{\citenamefont{Alhassid}(2000)}]{Alhassid00RMP}
\bibinfo{author}{\bibfnamefont{Y.}~\bibnamefont{Alhassid}},
  \bibinfo{journal}{Rev. Mod. Phys.} \textbf{\bibinfo{volume}{72}},
  \bibinfo{pages}{895} (\bibinfo{year}{2000}).

\bibitem[{\citenamefont{Sivan et~al.}(1996)\citenamefont{Sivan, Berkovits,
  Aloni, Prus, Auerbach, and Ben-Yoseph}}]{Sivan96}
\bibinfo{author}{\bibfnamefont{U.}~\bibnamefont{Sivan}},
  \bibinfo{author}{\bibfnamefont{R.}~\bibnamefont{Berkovits}},
  \bibinfo{author}{\bibfnamefont{Y.}~\bibnamefont{Aloni}},
  \bibinfo{author}{\bibfnamefont{O.}~\bibnamefont{Prus}},
  \bibinfo{author}{\bibfnamefont{A.}~\bibnamefont{Auerbach}}, \bibnamefont{and}
  \bibinfo{author}{\bibfnamefont{G.}~\bibnamefont{Ben-Yoseph}},
  \bibinfo{journal}{Phys. Rev. Lett.} \textbf{\bibinfo{volume}{77}},
  \bibinfo{pages}{1123} (\bibinfo{year}{1996}).

\bibitem[{\citenamefont{Prus et~al.}(1996)\citenamefont{Prus, Auerbach, Aloni,
  Sivan, and Berkovits}}]{Prus96}
\bibinfo{author}{\bibfnamefont{O.}~\bibnamefont{Prus}},
  \bibinfo{author}{\bibfnamefont{A.}~\bibnamefont{Auerbach}},
  \bibinfo{author}{\bibfnamefont{Y.}~\bibnamefont{Aloni}},
  \bibinfo{author}{\bibfnamefont{U.}~\bibnamefont{Sivan}}, \bibnamefont{and}
  \bibinfo{author}{\bibfnamefont{R.}~\bibnamefont{Berkovits}},
  \bibinfo{journal}{Phys. Rev. B} \textbf{\bibinfo{volume}{54}},
  \bibinfo{pages}{14289} (\bibinfo{year}{1996}).

\bibitem[{\citenamefont{Berkovits}(1998)}]{Berkovits98}
\bibinfo{author}{\bibfnamefont{R.}~\bibnamefont{Berkovits}},
  \bibinfo{journal}{Phys. Rev. Lett.} \textbf{\bibinfo{volume}{81}},
  \bibinfo{pages}{2128} (\bibinfo{year}{1998}).

\bibitem[{\citenamefont{Cohen et~al.}(1999)\citenamefont{Cohen, Richter, and
  Berkovits}}]{Cohen99}
\bibinfo{author}{\bibfnamefont{A.}~\bibnamefont{Cohen}},
  \bibinfo{author}{\bibfnamefont{K.}~\bibnamefont{Richter}}, \bibnamefont{and}
  \bibinfo{author}{\bibfnamefont{R.}~\bibnamefont{Berkovits}},
  \bibinfo{journal}{Phys. Rev. B} \textbf{\bibinfo{volume}{60}},
  \bibinfo{pages}{2536} (\bibinfo{year}{1999}).

\bibitem[{\citenamefont{Walker et~al.}(1999)\citenamefont{Walker, Montambaux,
  and Gefen}}]{Walker99}
\bibinfo{author}{\bibfnamefont{P.~N.} \bibnamefont{Walker}},
  \bibinfo{author}{\bibfnamefont{G.}~\bibnamefont{Montambaux}},
  \bibnamefont{and} \bibinfo{author}{\bibfnamefont{Y.}~\bibnamefont{Gefen}},
  \bibinfo{journal}{Phys. Rev. B} \textbf{\bibinfo{volume}{60}},
  \bibinfo{pages}{2541} (\bibinfo{year}{1999}).

\bibitem[{\citenamefont{Levit and Orgad}(1999)}]{Levit99}
\bibinfo{author}{\bibfnamefont{S.}~\bibnamefont{Levit}} \bibnamefont{and}
  \bibinfo{author}{\bibfnamefont{D.}~\bibnamefont{Orgad}},
  \bibinfo{journal}{Phys. Rev. B} \textbf{\bibinfo{volume}{60}},
  \bibinfo{pages}{5549} (\bibinfo{year}{1999}).

\bibitem[{\citenamefont{Ahn et~al.}(1999)\citenamefont{Ahn, Richter, and
  Lee}}]{Ahn99}
\bibinfo{author}{\bibfnamefont{K.}~\bibnamefont{Ahn}},
  \bibinfo{author}{\bibfnamefont{K.}~\bibnamefont{Richter}}, \bibnamefont{and}
  \bibinfo{author}{\bibfnamefont{I.}~\bibnamefont{Lee}},
  \bibinfo{journal}{Phys. Rev. Lett.} \textbf{\bibinfo{volume}{83}},
  \bibinfo{pages}{4144} (\bibinfo{year}{1999}).

\bibitem[{\citenamefont{Bonci and Berkovits}(1999)}]{Bonci99}
\bibinfo{author}{\bibfnamefont{L.}~\bibnamefont{Bonci}} \bibnamefont{and}
  \bibinfo{author}{\bibfnamefont{R.}~\bibnamefont{Berkovits}},
  \bibinfo{journal}{Europhys. Lett.} \textbf{\bibinfo{volume}{47}},
  \bibinfo{pages}{708} (\bibinfo{year}{1999}).

\bibitem[{\citenamefont{Stopa}(1996)}]{Stopa96}
\bibinfo{author}{\bibfnamefont{M.}~\bibnamefont{Stopa}},
  \bibinfo{journal}{Phys. Rev. B} \textbf{\bibinfo{volume}{54}},
  \bibinfo{pages}{13767} (\bibinfo{year}{1996}).

\bibitem[{\citenamefont{Hirose and Wingreen}(2002)}]{Hirose02}
\bibinfo{author}{\bibfnamefont{K.}~\bibnamefont{Hirose}} \bibnamefont{and}
  \bibinfo{author}{\bibfnamefont{N.~S.} \bibnamefont{Wingreen}},
  \bibinfo{journal}{Phys. Rev. B} \textbf{\bibinfo{volume}{65}},
  \bibinfo{pages}{193305} (\bibinfo{year}{2002}).

\bibitem[{\citenamefont{Blanter et~al.}(1997)\citenamefont{Blanter, Mirlin, and
  Muzykantskii}}]{Blanter97}
\bibinfo{author}{\bibfnamefont{Y.~M.} \bibnamefont{Blanter}},
  \bibinfo{author}{\bibfnamefont{A.~D.} \bibnamefont{Mirlin}},
  \bibnamefont{and} \bibinfo{author}{\bibfnamefont{B.~A.}
  \bibnamefont{Muzykantskii}}, \bibinfo{journal}{Phys. Rev. Lett.}
  \textbf{\bibinfo{volume}{78}}, \bibinfo{pages}{2449} (\bibinfo{year}{1997}).

\bibitem[{\citenamefont{Ullmo and Baranger}(2001)}]{Ullmo01b}
\bibinfo{author}{\bibfnamefont{D.}~\bibnamefont{Ullmo}} \bibnamefont{and}
  \bibinfo{author}{\bibfnamefont{H.~U.} \bibnamefont{Baranger}},
  \bibinfo{journal}{Phys. Rev. B} \textbf{\bibinfo{volume}{64}},
  \bibinfo{pages}{245324} (\bibinfo{year}{2001}).

\bibitem[{\citenamefont{Usaj and Baranger}(2001)}]{Usaj01}
\bibinfo{author}{\bibfnamefont{G.}~\bibnamefont{Usaj}} \bibnamefont{and}
  \bibinfo{author}{\bibfnamefont{H.~U.} \bibnamefont{Baranger}},
  \bibinfo{journal}{Phys. Rev. B} \textbf{\bibinfo{volume}{64}},
  \bibinfo{pages}{201319} (\bibinfo{year}{2001}).

\bibitem[{\citenamefont{Usaj and Baranger}(2002)}]{Usaj02}
\bibinfo{author}{\bibfnamefont{G.}~\bibnamefont{Usaj}} \bibnamefont{and}
  \bibinfo{author}{\bibfnamefont{H.~U.} \bibnamefont{Baranger}}
  (\bibinfo{year}{2002}), \bibinfo{note}{arXiv:cond-mat/0203074}.

\bibitem[{\citenamefont{Aleiner et~al.}(2002)\citenamefont{Aleiner, Brouwer,
  and Glazman}}]{Aleiner02}
\bibinfo{author}{\bibfnamefont{I.~L.} \bibnamefont{Aleiner}},
  \bibinfo{author}{\bibfnamefont{P.~W.} \bibnamefont{Brouwer}},
  \bibnamefont{and} \bibinfo{author}{\bibfnamefont{L.~I.}
  \bibnamefont{Glazman}}, \bibinfo{journal}{Phys. Rep.}
  \textbf{\bibinfo{volume}{358}}, \bibinfo{pages}{309} (\bibinfo{year}{2002}).

\bibitem[{\citenamefont{Bohigas}(1990)}]{Bohigas90}
\bibinfo{author}{\bibfnamefont{O.}~\bibnamefont{Bohigas}}, in
  \emph{\bibinfo{booktitle}{Chaos and Quantum Physics}}, edited by
  \bibinfo{editor}{\bibfnamefont{M.~J.} \bibnamefont{Giannoni}},
  \bibinfo{editor}{\bibfnamefont{A.}~\bibnamefont{Voros}}, \bibnamefont{and}
  \bibinfo{editor}{\bibfnamefont{J.}~\bibnamefont{Jinn-Justin}}
  (\bibinfo{publisher}{North-Holland}, \bibinfo{address}{Amsterdam},
  \bibinfo{year}{1990}), pp. \bibinfo{pages}{87--199}.

\bibitem[{\citenamefont{Brouwer et~al.}(1999)\citenamefont{Brouwer, Oreg, and
  Halperin}}]{Brouwer99}
\bibinfo{author}{\bibfnamefont{P.~W.} \bibnamefont{Brouwer}},
  \bibinfo{author}{\bibfnamefont{Y.}~\bibnamefont{Oreg}}, \bibnamefont{and}
  \bibinfo{author}{\bibfnamefont{B.~I.} \bibnamefont{Halperin}},
  \bibinfo{journal}{Phys. Rev. B} \textbf{\bibinfo{volume}{60}},
  \bibinfo{pages}{R13 977} (\bibinfo{year}{1999}).

\bibitem[{\citenamefont{Kurland et~al.}(2000)\citenamefont{Kurland, Aleiner,
  and Altshuler}}]{Kurland00}
\bibinfo{author}{\bibfnamefont{I.~L.} \bibnamefont{Kurland}},
  \bibinfo{author}{\bibfnamefont{I.~L.} \bibnamefont{Aleiner}},
  \bibnamefont{and} \bibinfo{author}{\bibfnamefont{B.~L.}
  \bibnamefont{Altshuler}}, \bibinfo{journal}{Phys. Rev. B}
  \textbf{\bibinfo{volume}{62}}, \bibinfo{pages}{14886} (\bibinfo{year}{2000}).

\bibitem[{\citenamefont{Jacquod and Stone}(2001)}]{Jacquod01}
\bibinfo{author}{\bibfnamefont{P.}~\bibnamefont{Jacquod}} \bibnamefont{and}
  \bibinfo{author}{\bibfnamefont{A.~D.} \bibnamefont{Stone}},
  \bibinfo{journal}{Phys. Rev. B} \textbf{\bibinfo{volume}{64}},
  \bibinfo{pages}{214416} (\bibinfo{year}{2001}).

\bibitem[{\citenamefont{Oreg et~al.}(2001)\citenamefont{Oreg, Brouwer, Waintal,
  and Halperin}}]{Oreg01}
\bibinfo{author}{\bibfnamefont{Y.}~\bibnamefont{Oreg}},
  \bibinfo{author}{\bibfnamefont{P.~W.} \bibnamefont{Brouwer}},
  \bibinfo{author}{\bibfnamefont{X.}~\bibnamefont{Waintal}}, \bibnamefont{and}
  \bibinfo{author}{\bibfnamefont{B.~I.} \bibnamefont{Halperin}}
  (\bibinfo{year}{2001}), \bibinfo{note}{arXiv:cond-mat/0109541}.

\bibitem[{\citenamefont{Simmel et~al.}(1997)\citenamefont{Simmel, Heinzel, and
  Wharam}}]{Simmel97}
\bibinfo{author}{\bibfnamefont{F.}~\bibnamefont{Simmel}},
  \bibinfo{author}{\bibfnamefont{T.}~\bibnamefont{Heinzel}}, \bibnamefont{and}
  \bibinfo{author}{\bibfnamefont{D.~A.} \bibnamefont{Wharam}},
  \bibinfo{journal}{Europhys. Lett.} \textbf{\bibinfo{volume}{38}},
  \bibinfo{pages}{123} (\bibinfo{year}{1997}).

\bibitem[{\citenamefont{Patel et~al.}(1998)\citenamefont{Patel, Cronenwett,
  Stewart, Huibers, Marcus, Duru{\"o}z, Harris, Campman, and
  Gossard}}]{Patel98}
\bibinfo{author}{\bibfnamefont{S.~R.} \bibnamefont{Patel}},
  \bibinfo{author}{\bibfnamefont{S.~M.} \bibnamefont{Cronenwett}},
  \bibinfo{author}{\bibfnamefont{D.~R.} \bibnamefont{Stewart}},
  \bibinfo{author}{\bibfnamefont{A.~G.} \bibnamefont{Huibers}},
  \bibinfo{author}{\bibfnamefont{C.~M.} \bibnamefont{Marcus}},
  \bibinfo{author}{\bibfnamefont{C.~I.} \bibnamefont{Duru{\"o}z}},
  \bibinfo{author}{\bibfnamefont{J.~S.} \bibnamefont{Harris}},
  \bibinfo{author}{\bibfnamefont{J.~K.} \bibnamefont{Campman}},
  \bibnamefont{and} \bibinfo{author}{\bibfnamefont{A.~C.}
  \bibnamefont{Gossard}}, \bibinfo{journal}{Phys. Rev. Lett.}
  \textbf{\bibinfo{volume}{80}}, \bibinfo{pages}{4522} (\bibinfo{year}{1998}).

\bibitem[{\citenamefont{L{\"u}scher et~al.}(2001)\citenamefont{L{\"u}scher,
  Heinzel, Ensslin, Wegscheider, and Bichler}}]{Luscher01}
\bibinfo{author}{\bibfnamefont{S.}~\bibnamefont{L{\"u}scher}},
  \bibinfo{author}{\bibfnamefont{T.}~\bibnamefont{Heinzel}},
  \bibinfo{author}{\bibfnamefont{K.}~\bibnamefont{Ensslin}},
  \bibinfo{author}{\bibfnamefont{W.}~\bibnamefont{Wegscheider}},
  \bibnamefont{and} \bibinfo{author}{\bibfnamefont{M.}~\bibnamefont{Bichler}},
  \bibinfo{journal}{Phys. Rev. Lett.} \textbf{\bibinfo{volume}{86}},
  \bibinfo{pages}{2118} (\bibinfo{year}{2001}).

\bibitem[{\citenamefont{Ong et~al.}(2001)\citenamefont{Ong, Baranger, Higdon,
  Patel, and Marcus}}]{Ong01}
\bibinfo{author}{\bibfnamefont{T.~T.} \bibnamefont{Ong}},
  \bibinfo{author}{\bibfnamefont{H.~U.} \bibnamefont{Baranger}},
  \bibinfo{author}{\bibfnamefont{D.~M.} \bibnamefont{Higdon}},
  \bibinfo{author}{\bibfnamefont{S.~R.} \bibnamefont{Patel}}, \bibnamefont{and}
  \bibinfo{author}{\bibfnamefont{C.~M.} \bibnamefont{Marcus}}
  (\bibinfo{year}{2001}), \bibinfo{note}{unpublished}.

\bibitem[{\citenamefont{Parr and Yang}(1989)}]{ParrYang89}
\bibinfo{author}{\bibfnamefont{R.~G.} \bibnamefont{Parr}} \bibnamefont{and}
  \bibinfo{author}{\bibfnamefont{W.}~\bibnamefont{Yang}},
  \emph{\bibinfo{title}{Density-Functional Theory of Atoms and Molecules}}
  (\bibinfo{publisher}{Oxford University Press}, \bibinfo{address}{New York},
  \bibinfo{year}{1989}).

\bibitem[{\citenamefont{Tanatar and Ceperley}(1989)}]{Tanatar89}
\bibinfo{author}{\bibfnamefont{B.}~\bibnamefont{Tanatar}} \bibnamefont{and}
  \bibinfo{author}{\bibfnamefont{D.~M.} \bibnamefont{Ceperley}},
  \bibinfo{journal}{Phys. Rev. B} \textbf{\bibinfo{volume}{39}},
  \bibinfo{pages}{5005} (\bibinfo{year}{1989}).

\bibitem[{\citenamefont{Egger et~al.}(1999)\citenamefont{Egger, H{\"a}usler,
  Mak, and Grabert}}]{Egger99}
\bibinfo{author}{\bibfnamefont{R.}~\bibnamefont{Egger}},
  \bibinfo{author}{\bibfnamefont{W.}~\bibnamefont{H{\"a}usler}},
  \bibinfo{author}{\bibfnamefont{C.~H.} \bibnamefont{Mak}}, \bibnamefont{and}
  \bibinfo{author}{\bibfnamefont{H.}~\bibnamefont{Grabert}},
  \bibinfo{journal}{Phys. Rev. Lett.} \textbf{\bibinfo{volume}{82}},
  \bibinfo{pages}{3320} (\bibinfo{year}{1999}).

\bibitem[{\citenamefont{Pederiva et~al.}(2000)\citenamefont{Pederiva, Umrigar,
  and Lipparini}}]{Pederiva00}
\bibinfo{author}{\bibfnamefont{F.}~\bibnamefont{Pederiva}},
  \bibinfo{author}{\bibfnamefont{C.~J.} \bibnamefont{Umrigar}},
  \bibnamefont{and}
  \bibinfo{author}{\bibfnamefont{E.}~\bibnamefont{Lipparini}},
  \bibinfo{journal}{Phys. Rev. B} \textbf{\bibinfo{volume}{62}},
  \bibinfo{pages}{8120} (\bibinfo{year}{2000}).

\bibitem[{\citenamefont{Bohigas et~al.}(1993)\citenamefont{Bohigas, Tomsovic,
  and Ullmo}}]{Bohigas93}
\bibinfo{author}{\bibfnamefont{O.}~\bibnamefont{Bohigas}},
  \bibinfo{author}{\bibfnamefont{S.}~\bibnamefont{Tomsovic}}, \bibnamefont{and}
  \bibinfo{author}{\bibfnamefont{D.}~\bibnamefont{Ullmo}},
  \bibinfo{journal}{Phys. Rep.} \textbf{\bibinfo{volume}{223}},
  \bibinfo{pages}{43} (\bibinfo{year}{1993}).

\bibitem[{\citenamefont{Ullah}(1964)}]{Ullah64}
\bibinfo{author}{\bibfnamefont{N.}~\bibnamefont{Ullah}},
  \bibinfo{journal}{Nucl. Phys.} \textbf{\bibinfo{volume}{58}},
  \bibinfo{pages}{65} (\bibinfo{year}{1964}).

\bibitem[{\citenamefont{Folk et~al.}(2001)\citenamefont{Folk, Marcus,
  Berkovits, Kurland, Aleiner, and Altshuler}}]{Folk01}
\bibinfo{author}{\bibfnamefont{J.~A.} \bibnamefont{Folk}},
  \bibinfo{author}{\bibfnamefont{C.~M.} \bibnamefont{Marcus}},
  \bibinfo{author}{\bibfnamefont{R.}~\bibnamefont{Berkovits}},
  \bibinfo{author}{\bibfnamefont{I.~L.} \bibnamefont{Kurland}},
  \bibinfo{author}{\bibfnamefont{I.~L.} \bibnamefont{Aleiner}},
  \bibnamefont{and} \bibinfo{author}{\bibfnamefont{B.~L.}
  \bibnamefont{Altshuler}}, \bibinfo{journal}{Phys. Scripta}
  \textbf{\bibinfo{volume}{T90}}, \bibinfo{pages}{26} (\bibinfo{year}{2001}).

\bibitem[{\citenamefont{Payne et~al.}(1992)\citenamefont{Payne, Teter, Allan,
  Arias, and Joannopoulos}}]{Payne92}
\bibinfo{author}{\bibfnamefont{M.~C.} \bibnamefont{Payne}},
  \bibinfo{author}{\bibfnamefont{M.~P.} \bibnamefont{Teter}},
  \bibinfo{author}{\bibfnamefont{D.~C.} \bibnamefont{Allan}},
  \bibinfo{author}{\bibfnamefont{T.~A.} \bibnamefont{Arias}}, \bibnamefont{and}
  \bibinfo{author}{\bibfnamefont{J.~D.} \bibnamefont{Joannopoulos}},
  \bibinfo{journal}{Rev. Mod. Phys.} \textbf{\bibinfo{volume}{64}},
  \bibinfo{pages}{1045} (\bibinfo{year}{1992}).

\end{thebibliography}
\end{document}